\documentclass[11pt,twoside]{article}
\usepackage{asp2004}
\usepackage{psfig}
\usepackage{epsf}
\usepackage{graphics}
\usepackage{lscape}
\def\ee #1   {\times 10^{#1}}          % \ee p       10^p
\def\ut #1 #2{ \, \mathrm{#1}^{#2}} % \ut unit p  unit^p
\def\u #1    { \, \mathrm{#1}}          % \u unit     unit
\def\msol    {\hbox{$M_\odot$}}
\def\lsol    {\hbox{$L_\odot$}}

\def\etal    {{\it et al. }}                     % et al

\def\arg#1{{\it#1\/}}

\def\edcomment#1{\iffalse\marginpar{\raggedright\sl#1\/}\else\relax\fi}
\reversemarginpar

\begin{document}
\title{Young Stellar  Clusters, WR-type Phenomenon and the Origin of  the
Galactic Center 
Nonthermal Radio Filaments}
\author{F. Yusef-Zadeh}
\affil{Dept.  Physics and Astronomy,
Northwestern University,  Evanston, IL.
60208}

\begin{abstract}

Recent observations of the Arches cluster located within a projected distance of 30 pc from the dynamical center of the Galaxy have
shown the presence of diffuse and discrete X-ray continuum emission, diffuse 6.4keV line emission as well as thermal and nonthermal
radio continuum emission. This young and dense stellar cluster is also recognized to be within the 95\% error circle of an identified
steady source of $\gamma$-ray emission associated with the EGRET source 3EG J1746--2851. Much of the thermal
 and nonthermal emission can be explained by shocked gas resulting from
colliding winds originating from massive binaries within the cluster.  In
particular, we argue that nonthermal particles could upscatter the
radiation field of the cluster by ICS and account for the $\gamma$-ray
emission. We also consider that the fluorescent 6.4 keV line emission may
be the result of the impact of low-energy relativistic particles on 
neutral gas distributed in the vicinity of the cluster. Lastly, we sketch
an interpretation in which young stellar clusters and massive young binary
systems
are responsible for the
origin of nonthermal radio filaments found throughout the inner 300pc of
the Galaxy. The collimation   of the   nonthermal
filaments  may be  done in the colliding wind region by  the ionized surface of 
individual mass-losing  stars of massive binary systems. 
In this picture,  a  WR-type phenomenon is expected to 
power a central star burst in the Galactic center in order to account for 
all the observed filaments.   

\end{abstract}
\thispagestyle{plain}

\section{Introduction}

One of the most important results of recent studies of the Galactic center
is the revelation of a large concentration of dark matter which is
coincident with the bright compact radio source Sgr A* (Sch\"odel et al.
2003; Ghez et al. 2003). Another important result is the discovery of
three clusters of young, massive stars within the projected distance of 30
pc from the Galactic center.  These objects are not run-of-the-mill
clusters in the Galaxy and it is remarkable that three such young and
compact systems are found in a small volume of the Galaxy. This region is 
known for its high 
extinction and source confusion and  star formation  
has to overcome the strong tidal effect in this high pressure environment.
These high density
stellar clusters consist of mainly O and WR stars with individual stellar
masses greater than 20 \msol as their winds should affect their
surrounding interstellar environment. Here, we focus on the Arches
cluster, the densest and possibly the youngest of these systems.  
This cluster provides an excellent laboratory to study thermal and
nonthermal processes operating  in a small volume. The other two clusters
are
the Quintuplet and the central Sgr A clusters, as discussed by D. Figer
and A.  Ghez in these  proceedings.

The young clusters such as the Arches cluster can be   identified not
only by  near-IR wavelength technique but also in X-rays and radio
wavelengths. 
In X-ray regime, clusters can be detected by their hot thermal
X-ray emission due to wind-wind collision in binary
systems whereas in high frequency radio regime, a cluster of free-free 
emitting sources are identified with ionized stellar winds. 
Nonthermal radio emission at low frequencies may be another method of 
detecting dense stellar clusters.  
The production of relativistic particles in
dense stellar clusters as well as OB-WR binary systems have important
consequences in the chemistry of interstellar clouds in the vicinity of
the cluster. Here we argue that dense stellar clusters and WR-OB binary
systems are seeds of relativistic particles that can be responsible for
production of nonthermal radio filaments observed throughout the Galactic
center.

\subsection{The Arches  Cluster G0.12+0.02}
\subsubsection{\bf X-ray Emission}

The Arches cluster consists of about 150 O star candidates with stellar
masses greater than 20~M$_\odot$ (Figer et al. 1999).  This cluster is
$\sim15''$ across, with an estimated density of $3\times10^5$ \msol
pc$^{-3}$ within its inner 9$''$ (0.36 pc)  (Cotera et al. 1996; Serabyn,
Shupe \& Figer 1998).  It is the densest known stellar cluster in the
Galaxy, denser than R136, the central cluster of 30 Dor in the LMC.  The
age of the Arches cluster is estimated to be 1-2 Myrs (e.g.  Figer \etal
1999). X-ray emission from the Arches cluster has recently been detected
using {\it Chandra} observations (Yusef-Zadeh \etal 2002; Law and
Yusef-Zadeh 2004).  These observations identify two bright compact sources
A1S, A1N lying at the core of the cluster and one source A2 lying at the
boundary of the cluster about 10$''$ away from the core.  These sources
coincide with mass-losing WN/Of stars. The spectral analysis of individual
sources A1N, A1S and A2 give fit values of temperatures $<$ 2.3 keV. The
spectra of the sources from the inner 15$''$ of the core may be fit with a
two-temperature absorbed model (Yusef-Zadeh et al. 2002) or a single
temperature and a power law (Law and Yusef-Zadeh 2004).  The quality of
each fit is essentially the same, as measured by $\chi^2$ statistics,
however, the unabsorbed flux and the photon index from the powe-rlaw
component are very difficult to constrain. The unabsorbed X-ray luminosity
from the core of the cluster ranges between 0.5 and 3$\times 10^{35}$ erg
s$^{-1}$ depending on which spectral fits have been used.

One of the most interesting result coming from  X-ray observations of the
Arches
cluster is the detection of extended emission surrounding the cluster.
This large scale diffuse ovoid feature (A3) with dimensions of
approximately 90$''\times 60''$ (3.6 $\times$ 2.4 pc) extends well beyond
the edge of the cluster, which is $<$ 15$''$ in diameter. 
 Theoretical work
studying the nature of X-ray emission from the collision of stellar winds
in a dense cluster environment such as the Arches cluster predicts a
cluster wind escaping from the outer boundary of the cluster (Cant\'o
\etal 2000, Raga et al. 2001).  These authors consider that 
shocked gas arising
from stellar winds is in the form of discrete X-ray sources as well as
diffuse X-ray emission, the so-called ``cluster wind".
More
recently, Rockefeller et al. (2004) have calculated simulations of X-ray
luminosities from the Arches cluster where both diffuse and compact X-ray
sources have been accounted for by wind sources with a varying
degree of mass-loss
rates. It is possible that the continuum extended emission is due
to the ``cluster wind''  as predicted theoretically. 

 Extended
6.4 keV line component is also detected. 
The spectrum of the extended source A3  can be fitted by a single thermal
bremsstrahlung and an
additional
Gaussian contributed by fluorescent Fe K$\alpha$ 6.4 keV line.  
The nature of diffuse 6.4 keV emission is not clear.  This diffuse
component associated with A3 may be produced by the 
scattering  of radiation from an adjacent  molecular cloud. 
The irradiation of the cloud 
may  also contaminate the "cluster wind" emission which is expected to
be extended. 
 The true distribution of  
scattered radiation and the cluster wind emission is a difficult task 
that should be done in future sensitive measurements. 
The 
source of hard X-ray emission responsible for irradiation of the cloud  
could be the Arches cluster itself 
or perhaps Sgr A$^*$ assuming that it has been active in the past. 
An alternative mdel is that the impact of low-energy cosmic
rays produces  the 6.4 keV line and  continuum emisison from A3. 
The relativistic particles  needed for this mechanism could arise from the
wind-wind collsion from the cluster itself, as discussed below.
It has recently  
been demonstrated that bremsstrahlung from low-energy
cosmic-ray electrons can  naturally explain the spectrum below 10 keV
as well as  the strength of the Fe 6.4 keV line emission (Valinia \etal
2000).  This model has been applied successfully to a dense molecular 
cloud G0.13-0.13 near the Galactic center (Yusef-Zadeh et al. 2003). 

\subsubsection{\bf Radio  Emission}

Radio continuum emission from a cluster of nine stellar sources AR1-9 has
recently been
detected toward the Arches cluster (Lang et al. 2001).  The radio spectra
and near-IR spectral type
of the cluster of eight radio stars are consistent with
ionized stellar
winds arising from mass-losing WN and/or Of stars with mass-loss rates
ranging between $3\times10^{-5}$ to $1.7\times10^{-4}$ \msol yr$^{-1}$
(Lang \etal 2001).  Free-free emission resulting from ionized stellar
outflow has a positive spectral index (F$_{\nu}\propto\nu^{0.6}$; Panagia
\& Felli 1975). With the exception of AR6, all sources show a spectrum
ranging between $\alpha$=0.3 and 0.9.  The X-ray sources A1N and A1S at
the
core of the cluster coincide with ionized stellar wind sources AR1 and
AR4, respectively (Lang et al.\ 2001). Interestingly, both these bright
early type  stellar sources are variable in radio wavelengths (Lang 2003).
 
More recently, radio
observations of the Arches cluster indicate nonthermal emission from this
cluster at 327 MHz (Yusef-Zadeh et al. 2003). The high frequency radio
emission from the cluster is compact and arises from mass-losing stellar
sources, whereas the low-frequency radio emission appears to be diffuse
with nonthermal characteristics. The evidence for nonthermal emission 
at radio wavelengths strengthens the single temperature and power-law fit
model to the X-ray emission from the inner 15$''$ of the Arches cluster
as discussed above.

\subsubsection{\bf $\gamma$-ray  Emission}

The existence of nonthermal particles within the core of a luminous young
stellar cluster suggests the possibility that the nonthermal X-ray or
$\gamma$-ray emission could result from upscattering of the radiation
field with a luminosity of $10^{7.8}\lsol$ by nonthermal particles
(Ozernoy, Genzel \& Usov 1997; Yusef-Zadeh et al. 2003).  In fact, the
Arches cluster is displaced only by $\approx200''$ from the nominal
position of the unidentified EGRET source 3EG J1746--2851 (Hartman et al.
1999), located well within the 95\% error radius of $0.13^0$.  This steady
and strong $\gamma$-ray source has a photon index 1.7 and a flux
of 1.2$\times10^{-6}$ photons cm$^{-2}$ s$^{-1}$ with energies greater
than
100 MeV. The $E^{-1.7}$ photon spectrum of 3EG J1746--2851 could be
produced by inverse Compton scattering from a distribution of relativistic
electrons. The spectral index of
$\alpha$=0.7 is  consistent within the
uncertainty of  the spectral index
value of nonthermal radio emission from the Arches cluster (Yusef-Zadeh,
Law and Wardle 2003).

\subsection{Discussion}
\subsubsection{\bf Colliding Wind Binaries}

Radio and X-ray observations of the Arches cluster provide  the evidence
for interacting binary systems within the cluster. The strongest argument
in favor of such a suggestion comes from the variability of radio emission
from stellar sources at high frequencies (Lang 2003).  The variable radio
sources AR1 and AR4 coincide with the peak of X-ray emission A1S and A1N
at the core of the cluster.  Radio emission from binary stars could vary
between fully thermal and nonthermal spectrum and many WR-OB binary
systems have displayed  this characteristic (e.g., Chapman et al. 1999).
Additional support comes from the nonthermal spectrum of AR6 and the
nonthermal extended emission from the cluster, both of which are unlikely
to be produced by single stars.  Yet another reason for the suggestion of
binary systems within the cluster comes from high X-ray luminosity of
discrete sources as well as a high ratio of X-ray to radio flux for
individual sources. These ratios are more consistent with those of known
WR-OB binary systems (Chapman et al. 1999)

\subsubsection{\bf Nonthermal  Radio Emission}

One of the most intriguing aspects of radiative properties of the Arches
cluster is the detection of low-frequency nonthermal emission at 327 MHz.
One of the candidates for production of nonthermal emission is colliding
wind binaries within the cluster. However, free-free absorption in these
sources will not allow nonthermal radiation to escape from the photosphere
radius of individual members of WR-OB stars with optical depth of 1 at 327
MHz.  For example, the photospheric radius of an O star wind is more than
10$^4 R_\odot$ which is as large as the size of the binary system. Other
effects that are also important to suppress the emission at low frequencies
are synchrotron self-absorption and Razin effect (Dougherty et al. 2003;
Pittard in  these proceedings).  The fact that nonthermal low-frequency
emission has been detected  suggests that neither of these effects may
be important. Deatiled theoretical work by Dougherty et al. (2003) 
suggests that the ratio of relativistic to thermal energy density 
is critical  for suppression of low-frequency emission due 
to synchrotron self absorption and Razin effects.  However, if the
the relativistic energy density  as well as  the maximum energy of
relativistic particles $\gamma$ are  
high enough, neither Razin nor synchroreon self-absorption can be
important. This implies that  the acceleration of particles, whatever the
mechanism (e.g., Fermi acceleration, reconnection) is  very efficient
in these systems. 
Alternatively, the
nonthermal emission may
arise from the contibution of the  cluster wind being shocked at the
interface between the
stellar cluster and the dense ISM. Another contribution  to
explain the
origin of low-frequency radio emission is that the emission arises from  
the colliding winds between individual
members of the cluster.
The average separation of stars  within the cluster
is larger than  the O star photospheric
radius of
optical
depth 1 at 327 MHz.  The evidence for diffuse nonthermal emission at 327
MHz from the Arches cluster is not inconsistent with this picture.

\subsubsection{\bf Nonthermal Radio Filaments}

Over the last two decades a large number of radio observations have shown
the existence of dozens of systems of filamentary structures within the
inner two degrees of the Galactic center (e.g., Nord et al. 2004;  
Yusef-Zadeh, Hewitt and Cotton 2004).  Their transverse filament
dimensions are 0.1 to 1pc (2$''$ to 20$''$ at the distance of the Galactic
center) and their length is on  the order of tens of parsecs.  Their
strongly linearly polarized emission plus their radio spectral index
distribution suggest a nonthermal synchrotron origin.  
Recent observations indicate that  the longest and most prominent 
filaments run roughly perpendicular to the Galactic plane ($>$ 5$'$)
whereas 
the short
filaments do not show a preferred orientation perpendicular
to the Galactic plane.

One of the motivations for the hypothesis that nonthermal radio filaments
near the Galactic center originate from dense clusters of masssive stars
comes from recent radio surveys indicating that many of the radio
filaments are concentrated in the vicinity of star forming regions. The
other motivation is the revelation that nonthermal particles are easily
generated by colliding winds of massive young binaries. The idea of
stellar acceleration of particles to relativistic energies was originally
set forth by Rosner and Bodo (1996) in order to explain the origin of
nonthermal filaments.  Unlike many other models of the Galactic center
filaments, this idea was able to explain the transverse dimension of the
filaments by matching it with the size of the wind bubble created by winds
of mass-losing stars.  This idea was recently expanded 
by suggesting that nonthermal emission from the Galactic center filaments
originates
from the shocked region of the colliding stellar winds of young clusters
(Yusef-Zadeh 2003). The evidence for 
the generation of nonthermal particles
which are accelerated by  young massive binary systems is
overwhelming.  In fact, numerous WR-type stars are found in the three 
young clusters in the Galactic center region. 
Based on  a high fraction of massive stars found in binary
and
multiple systems, we provided  evidence that these systems are
also distributed within
the Arches cluster. More recently, the nonthermal
filaments have been  interpreted in terms of jets launched by embedded 
stars or
clusters as they extract mass and energy propagating in a dense ISM of the
Galactic center region (Yusef-Zadeh and K\"onigl 2004).

Theoretically, the key issue is to explain the highly collimated nature of
 nonthermal filaments in the context of a jet model launched from massive
binaries  and dense clusters. In this schematic model, we suggest that the
wind collision  region is the site of acceleration of relativistic
particles in a close binary system. We assume that the dense stellar
clusters are at their earliest phase of evolution when they are most
compact containing WR-OB binary systems whose members orbit each other 
with a short
period. The colliding wind region lies at a distance where the winds have
gained their  terminal speeds.  Stars within the binary system are so 
close to
each other that the surface of the ionized winds from individual stars
deviates from spherical geometry due to tidal effects.  In this picture,
the radial density distribution of ionized wind does not follw r$^{-2}$
and the colliding wind region has an angular dependence in the orbital
plane where mass-loss has an angular dependence (e.g., Friend and Castor
1982). The strongest synchrotron emissivity is expected to be along the
axis where the wind velocity and density are highest. In the case where
there is spherical symmetry, the intrinsics synchrotron emissivity from
the wind collision site depends on D$^{-1/2}$ where D is the binary
separation (Doughrty et al. 2003).  The relativistic particles generated
at this site is expected to be advected by the high pressure thermal gas
as they run perpendicular to the orbital plane and form jet-like
structures.  A recent study of SS443 suggests that WR-type phenomenon is
responsible for the collimated jet of well-known system (Fuchs in these
proceedings).  We propose that the distorted surface of the ionized winds
from individual WR stars, as shown schematically in Figure 1, 
 acts as a wall that collimates nonthermal gas the direction perpendicular 
to the orbital plane of a close
binary system.  We speculate that the most prominent nonthermal filaments
are expected to be associated with the early phase of very dense massive
clusters whereas the faint and short filaments are accounted for by
isolated massive binary systems.  In both scenarios, the filaments are
expected to be located in the vicinity of young stellar sources with X-ray
counterparts.  Due to strong effects of synchrotron self-absorption, Razin
and free-free absorption, radio emission from the region where the
filaments originate are expected to be weak.  If this picture is correct,
a large number of hidden WR-OB binaries and young clusters are required to
explain numerous filaments found in the Galactic center region.  Such a
large number of massive young stars and clusters in a small volume implies
a star burst activity in the nucleus of the Galaxy. This is not
inconsistent with recent ISO measurements of the excitation condition of
the Galactic center region suggesting a low-excitation star burst activity
(Rodriguez-Fernandez and Martin-Pintado 2004).  Additional support for a
powerful star burst activity in the last several million years comes from
large-scale Galactic winds in this region (Bland-Hawthorn and Cohen 2003).  
A more detailed account of this model will be given elsewhere.

\begin{quote}
{\bfseries Acknowledgments.} We thank A. K\"onigl and R.  Taam  for 
useful discussion.
\end{quote}

\section{References}

\begin{quote}

Bland-Hawthorn, J. \& Cohen, M. 2003, ApJ, 582, 246\\
Cant\'o, J., Raga, A.C. \& Rodriguez, L.F. 2000, ApJ, 536,  896\\
Chapman, J.M.  et al. 1999, ApJ, 518, 890\\  
Cotera, A. S., et al. 1996, ApJ, 461, 750\\
Dougherty, S.M. et al. 2003, A\&A, 217-233\\
Eichler, D. \& Usov, V. 1993, ApJ, 402, 271\\
Figer et al. 1999, ApJ, 514, 202\\
Friend, D.B. \& Castor, J.I.  1982, ApJ, 261, 263\\
Ghez, A. et al. 2003 ApJ, 586, L127\\
Hartman, R.C.  et al.:  1999, ApJS, 123, 79\\
Lang, C. 2003, NRAO Newsletter 95, 16 \\
Lang, C.C., Goss, W.M. \& Morris, M.  2001a, AJ, 121, 2681\\
Lang, C.C., Goss, W.M. \& Rodriguez, L.F.  2001b, ApJ, 551, L143\\   
Law, C.  and Yusef-Zadeh, F. 2004, ApJ, 611, 858\\
Nord, M. et al. 2004, ApJ, 611, 858\\
Ozernoy, L.M., Genzel, R. \& Usov, V.  1997, MNRAS, 288, 237\\
Panagia \& Felli 1975, A\&A, 35, 1\\
Raga, A.C.  et al. 2001, ApJ, 559, L33\\
%Rockefeller, G., Fryer, C., Melia, F. \& Wang, Q.D. 2004, ApJ,
Rockefeller, G. et al.  2004, ApJ, submitted\\
Rodriguez-Fernandez, N.J. \& Martin-Pintado, J. 2004, A\&A, in press\\
Rosner, R. \& Bodo, G. 1996, ApJ, 470, L49\\
Sch\"odel, R., Ott, T., Genzel, R. et al. 2003, ApJ, 596, 1015\\
Serabyn, E., Shupe, D., \& Figer, D. F. 1998, Nature, 394, 448\\
Valinia, A. et al.  2000, ApJ, 543, 733\\
Yusef-Zadeh, F. 2003, ApJ, 598, 325\\
Yusef-Zadeh \& K\"onigl 2004, in ``formation and evolution of massive
young star clusters'', eds: H.  Lamers et al. (astro-ph/0403201\\
Yusef-Zadeh et al. 2003, ApJ, 590, L103\\
Yusef-Zadeh et al. 2002, ApJ, 570, 665\\
Yusef-Zadeh, F., Hewitt, J. \& Cotton, W. 2004, ApJS, in press\\ 
Yusef-Zadeh, F., Morris, M. \& Chance D. 1984, Nature, 310, 557\\
\end{quote}

\begin{quote}
\begin{figure}[!ht]
\plotone{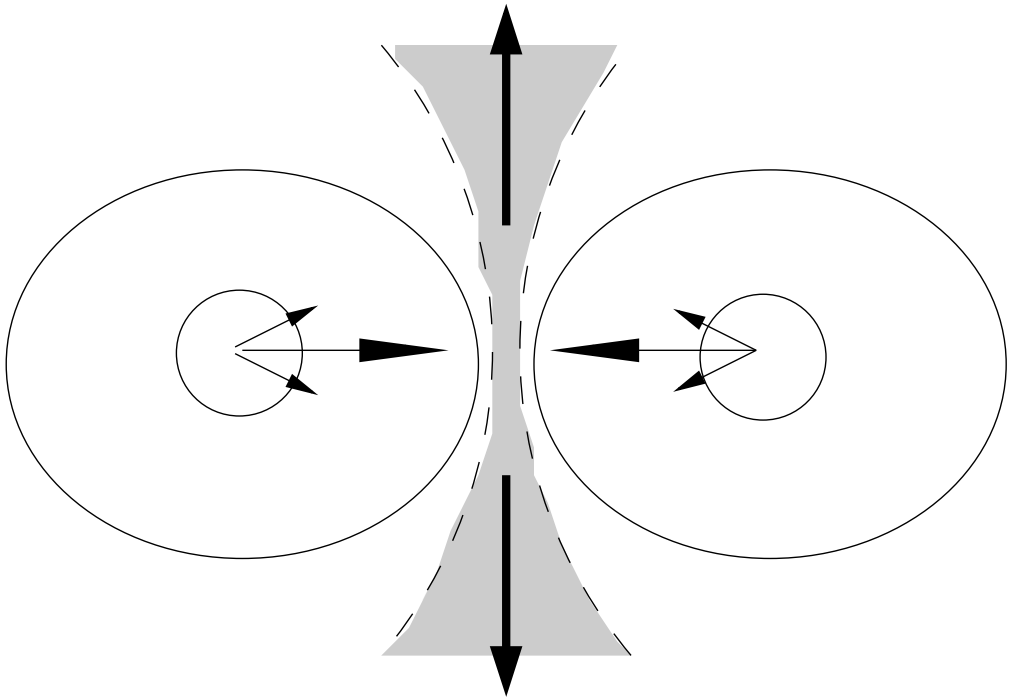}
\caption{A schematic diagram showing the geometry of the colliding wind 
region in a close binary system where the surface of ionized wind is asymmetric 
by tidal and centrifugal effects. For simplicity, both mass-losing
young stars are 
assumed to have the same spectral type. The outflow representing a  galactic
center filament consists of 
relativistic particles  produced by wind-wind collison and is collimated by the 
surface of the ionized winds from the mass-losing stars}
\end{figure}
\end{quote}

\end{document}